\documentclass[prd,preprintnumbers,
twocolumn,
eqsecnum,floatfix,letterpaper,superscriptaddress,nofootinbib]{revtex4}
\usepackage{mathtools}
\usepackage{multirow}
\usepackage{amsmath,amssymb,graphicx}
\usepackage[printwatermark]{xwatermark}
\usepackage{lipsum}
\usepackage{bm}
\usepackage{color}
\usepackage{booktabs}
\usepackage{tabularx}
\usepackage{times}
\usepackage{microtype}
\usepackage{booktabs}
\usepackage[caption=false]{subfig}

\usepackage{csquotes}
\usepackage{array}
\usepackage{verbatimbox}

\newcolumntype{M}[1]{>{\centering\arraybackslash}m{#1}}
\newcolumntype{N}{@{}m{0pt}@{}}

\usepackage[normalem]{ulem}
\usepackage[varg]{txfonts}
\usepackage{dirtytalk}
\usepackage[colorlinks, pdfborder={0 0 0}]{hyperref}
\definecolor{LinkColor}{rgb}{0.75 , 0, 0}
\definecolor{CiteColor}{rgb}{0, 0.5, 0.5}
\definecolor{UrlColor}{rgb}{0, 0, 0.75}
\hypersetup{linkcolor=LinkColor}
\hypersetup{citecolor=CiteColor}
\hypersetup{urlcolor=UrlColor}

\allowdisplaybreaks
\def\dphizero{\delta\hat{\phi}_{0}}
\def\dphitwo{\delta\hat{\phi}_{2}}
\def\dphithree{\delta\hat{\phi}_{3}}

\def\dphifivel{\delta\hat{\phi}_{5l}}

\newcommand{\mycomment}[1]{}
\def\msun{M_{\odot}}
\def\PCAone{\delta\hat{\phi}^{(1)}_{\mathrm{PCA}}}
\def\PCAtwo{\delta\hat{\phi}^{(2)}_{\mathrm{PCA}}}
\def\PCAthree{\delta\hat{\phi}^{(3)}_{\mathrm{PCA}}}
\def\PCAfour{\delta\hat{\phi}^{(4)}_{\mathrm{PCA}}}
\def\PCAfive{\delta\hat{\phi}^{(5)}_{\mathrm{PCA}}}

\begin{document}
\title{Enhancing the performance of multiparameter tests of general relativity with LISA using Principal Component Analysis}

\author{Sayantani Datta}\email{sdatta94@cmi.ac.in} 
\affiliation{Chennai Mathematical Institute, Siruseri, 603103, India}
\date{\today}

\begin{abstract}
The Laser Interferometer Space Antenna (LISA) will provide us with a unique opportunity to observe the early inspiral phase of supermassive binary black holes (SMBBHs) in the mass range of $10^5-10^6\,M_{\odot}$, that lasts for several years. 
It will also detect the merger and ringdown phases of these sources. Therefore, such sources are extremely useful for multiparameter tests of general relativity (GR), where parametrized deviations from GR at multiple post-Newtonian orders are simultaneously measured, thus allowing for a rigorous test of GR. However, 
the correlations of the deviation parameters with the intrinsic parameters of the system make
multiparameter tests extremely challenging to perform. We demonstrate the use of principal component analysis (PCA) to obtain a new set of deviation parameters, which are best-measured orthogonal linear combinations of the original deviation parameters. With the observation of an SMBBH of total redshifted mass, $\sim\mathrm{7\times10^5\,M_{\odot}}$ at a luminosity distance of 3 Gpc, we can estimate the five most dominant PCA parameters,
with 1-$\sigma$ statistical uncertainty of $\lesssim 0.2$. The two most dominant PCA parameters can be bounded to $\sim \mathcal{O}(10^{-4})$, while the third and fourth-dominant ones to $\sim \mathcal{O}(10^{-3})$. Measurement of the PCA parameters with such unprecedented precision with LISA makes them an excellent probe to test the overall PN structure of the GW phase evolution.
\end{abstract}
	
 \maketitle
\section{Introduction}
The monumental success of the Laser Interferometer Space Antenna (LISA) pathfinder~\cite{ALISA06} mission has ensured the technological feasibility of detecting gravitational waves (GWs) in the milli-Hertz band. It has paved the way for the European Space Agency's (ESA) L3 mission~\cite{LISA2017}, which is proposed to be launched in the mid-2030s. As per ESA's L3 mission, LISA will consist of three satellites connected by six laser links to form a triangular shape with an arm length of 2.5 million kilometers. It will orbit the Sun, following Earth's orbit around it. LISA is designed to operate in the millihertz regime ($\sim 10^{-4}$--$10^{-1}$ Hz) and will detect supermassive binary black holes (SMBBHs) with masses between $10^5$--$10^7\,M_{\odot}$, up to a redshift of z $\sim$ 20~\cite{LISA2017}, with signal to noise ratio (SNR) of $\gtrsim 100$~\cite{LISA2017}. It will observe the inspiral phase of the GWs from SMBBHs for several years when the black holes are far apart and spiral towards each other, driven by gravitational radiation reaction. It will also observe the merger and ringdown phases of the SMBBHs. All these factors make SMBBHs the {\it golden binaries} for LISA and ideal candidates to test the fundamental nature of gravity~\cite{GairLivRev}.
Einstein's theory of general relativity (GR) has passed all the tests that probe the weak-field regime~\cite{Will05LivRev}---the region of space-time where the gravitational field is weak and slowly varying. However, building a theory of gravity which conforms well to the quantum limit is a longstanding problem. It forms the primary motivation to test GR in the strong-field sector, where space-time is highly dynamic and nonlinear. The detection of gravitational waves~\cite{Discovery,GW151226,GW170104,GW170608,GW170814,GWTC1,Venumadhav:2019lyq,Zackay:2019btq,Abbott:2020niy} by terrestrial detectors like Advanced LIGO~\cite{aLIGO} and Virgo~\cite{aVirgo} during their first, second and third observing runs, have facilitated strong-field tests of GR~\cite{YunesSiemens2013,GairLivRev,SathyaSchutzLivRev09,Berti:2015itd,Arun:2013bp}. 
To date, the detected gravitational wave signals were found to be consistent with the predictions of GR~\cite{O1BBH,O2-TGR,Abbott:2020jks,LIGOScientific:2021sio,LIGOScientific:2018dkp}.
However, LISA will allow us to perform precision strong-field tests of GR with SMBBHs, for the first time~\cite{LISA2017,BCW05}.

{\it Parametrized tests of GR}~\cite{AIQS06a, AIQS06b, MAIS10, YunesPretorius09, LiEtal2011, TIGER, YYP2016} using GW signals from compact binary coalescences is a very important strong-field test of GR. Under this framework, one tests the consistency of the GW phase evolution in the inspiral regime with the predictions from Post-Newtonian approximation to GR, by measuring the coefficients at every post-Newtonian (PN) order (powers of v/c).
In the frequency domain, the phase evolution of a compact binary in quasi-circular orbits is written as~\cite{CF94},
\begin{equation}\label{eq:schematic-PN-phasing}
\Phi(f)=2\pi f\,t_c-\phi_c+\frac{3}{128\,\eta\,\mathrm{v}^5} \sum_{k=0}^N\left[
\big( \phi_k\,+\,\phi_{\rm k\,l}\,\ln \mathrm{v} \big) \,\mathrm{v}^k \right],
\end{equation}
where, $t_c$ and $\phi_c$ are the time and phase at the binary coalescence. The PN expansion parameter $\mathrm{v}$ is defined as $\mathrm{v}\equiv (\pi\, M\,f)^{1/3}$, where $M = m_1 + m_2$ is the total redshifted mass of the source. The total mass in the source frame is given by $M_{\rm source}=M/(1+z)$, where $z$ is the redshift of the source with respect to the detector.
The symmetric mass ratio is given as $\eta = q/(1+q)^2$, where $q=m_2/m_1$.
The PN coefficients, $\phi_k\,(k = 0, 2, 3, 4, 6, 7)$ and $\phi_{kl}\,(kl = 5, 6)$ are unique functions of the properties of the compact binary like their component masses and spins and encode information about various physical and nonlinear interactions~\cite{Bliving}. It is natural to expect that these PN coefficients may defer from GR 
in a modified theory of gravity due to modified GW generation and propagation~\cite{AlexanderYunes07a,AlexanderYunes07b,CSreview09}. Therefore, measuring the different PN coefficients and testing for consistency with their GR values will help us bound possible deviations from GR~\cite{AIQS06a,AIQS06b,Chamberlain:2017fjl,Barausse:2016eii}. The deviations from GR may be parametrized at every PN order by adding generic PN deformation parameters to the PN coefficients:
\begin{subequations}\label{eq:dev}
\begin{eqnarray} 
\phi_k  \rightarrow  \phi_k^{\rm GR}( 1+\delta\hat{ \phi}_k ),\\
\phi_{kl}  \rightarrow  \phi_{kl}^{\rm GR}( 1+\delta\hat{\phi}_{kl} ),
\end{eqnarray}
\end{subequations}
where, the PN deformation parameters, $\delta\hat{\phi}_k$ and $\delta\hat{ \phi}_{kl}$ are adimensional by construction. This constitutes a very useful {\it null} parametrization, as these phenomenological PN deformation parameters are absent in GR and have zero value but can have a non-zero value in a modified theory of gravity~\cite{Bliving,YYP2016}. 

A particular physical or nonlinear effect in GR may appear in multiple PN coefficients. For example, nonlinear effects like `tails', generated by scattering waves on the curvature associated with the Schwarzschild metric, appear at 1.5PN, 2.5PN, 3PN, and 3.5PN orders~\cite{BD88,BS93} in the GW phase evolution for nonspinning binaries. Physical effects like spin-orbit interactions appear in 1.5PN, 2.5PN, 3PN, and 3.5PN orders~\cite{KWWi93,BBuF06} and spin-spin effects in 2PN and 3PN orders in the GW phase.
As a result, a beyond-GR effect may be imprinted on multiple PN coefficients, leading to non-zero values for multiple PN deformation parameters. Hence, it is natural to {\it simultaneously} measure multiple PN deformation parameters, which is the essence of {\it multiparameter tests of GR}~\cite{AIQS06a,Gupta:2020lxa,Datta:2020vcj,Carl:multiparam}. However, due to the high correlations of the PN deformation parameters with the component masses and spins~\cite{AIQS06a,Gupta:2020lxa,Datta:2020vcj,Carl:multiparam}, multiparameter tests return uninformative posterior distributions on the PN deformation parameters.

In Ref.~\cite{Gupta:2020lxa,Datta:2020vcj}, it is shown that we can indeed measure all the PN deformation parameters simultaneously along with the GR parameters by using multiband observations of heavy stellar black holes with LISA in the milli-hertz band and next-generation terrestrial detectors like Cosmic Explorer (CE)~\cite{2019arXiv190704833R} and Einstein Telescope (ET)~\cite{Punturo:2010zz} in the kilo-hertz band. The full multiparameter test was possible because of the cancellation of correlations between various pairs of PN deformation parameters when information from the two complementary bands was added coherently~\cite{Datta:2020vcj}. However, this test can only be performed with a specific class of GW sources and requires data from both LISA and CE/ET. 
Multiparameter tests of GR using only ground-based detectors return unconstrained posterior distributions on the PN deformation parameters~\cite{methods-PCAO1O2}. 

Another approach toward performing multiparameter tests of GR is to construct a new set of null deformation parameters, which are linear combinations of the original PN deformation parameters and uncorrelated to each other so that they can be simultaneously measured. In Refs.~\cite{methods-PCAO1O2, Datta:2022izc}, we described the construction of this new set of deformation parameters using principal component analysis (PCA) on the variance-covariance matrix or its inverse---the Fisher matrix,  corresponding to the PN deformation parameters. The new orthogonal set of deformation parameters is unique for every source, and we will call them PCA parameters. The sub-dominant PCA parameters showed more sensitivity towards higher PN orders, capturing the late-time strong-field dynamics better than the dominant one. However, the information carried by the sub-dominant PCA parameters decreases steeply compared to the dominant one, making them hard to measure with a reasonable precision.

In this paper, we investigate the construction and measurability of the PCA parameters and their ability to probe higher PN order effects with the changing total mass of the SMBBH in the LISA band. We determine the mass range best suited for this test for which the most PCA parameters are measurable with the highest accuracy. We also compare the performance of the PCA parameters computed with LISA and next-generation detectors like CE and ET.

The paper is organized in the following way. In Sec.~\ref{sec:PCAmethod}, we describe the parameter space for multiparameter tests of GR, the method to construct the PCA parameters with LISA, and compute the uncertainties on the PCA parameters. In Sec.~\ref{sec:PCAerrors}, we find the total number of PCA parameters measurable with 1-$\sigma$ uncertainty below 20\%. We also study the variation in the measurement uncertainty as a function of the total mass of the SMBBHs. In Sec.~\ref{sec:PCAcomparison}, we compare the properties and performance of the PCA parameters computed with LISA, CE, and ET and summarize the results in Sec.~\ref{sec:conclusion}.
\section{Multiparameter tests of GR using principal component analysis}
\label{sec:PCAmethod}
\subsection{The parameter space for multiparameter tests of GR and the LISA noise PSD}
\label{sec:Parameterspace}
In the frequency domain, the GW strain from compact binary coalescences is schematically written as,
\begin{equation}\label{eq:GWwaveform}
{\tilde h}(f)={\cal A}(f) \, e^{i\Phi (f)},
\end{equation}
 where, $\Phi (f)$ is the phase of the gravitational wave whose structure in PN theory is shown in  Eq.~(\ref{eq:schematic-PN-phasing}). ${\cal A}(f)$ is the amplitude of the wave, which depends on the source sky location, polarization and inclination angle of the orbit and luminosity distance ($D_L)$ with respect to the detector, component masses ($m_1$ and $m_2$) and spins ($\chi_1$ and $\chi_2$) of the binary system. In the inspiral phase, ${\cal A}(f) \propto {\cal C} {\cal M}^{5/6}D_L^{-1}f^{-7/6}$, where ${\cal M}= M \eta^{3/5}$ is the chirp mass. ${\cal C}$ is a prefactor that contains the antenna response functions of the detector and is a function of the source location, polarization, and orbit's inclination angle.

We use the \textsc{IMRPhenomD}~\cite{Khan2016} waveform model, whose amplitude and phase contain the inspiral, merger, and ringdown stages of the compact binary evolution. The amplitude of the waveform contains only the leading order quadrupolar mode. The waveform model assumes that the spin angular momentum of the binary components is (anti-)aligned to the orbital angular momentum and hence does not have any precession. In order to model beyond-GR deviations, the fractional PN deformation parameters are introduced at every PN order in the inspiral phase of the waveform as shown in Eq.~(\ref{eq:dev}), except the non-logarithmic term at 2.5PN as the corresponding coefficient is frequency independent and is entirely degenerate with the phase at coalescence, $\phi_c$. Therefore, the parameter space ($\vec{\theta}$) is composed of eight PN deformation parameters ($\vec{\theta}_{\mathrm{NGR}}$) along with seven GR parameters ($\vec{\theta}_{\mathrm{GR}}$), which completely describe our parametrization:
\begin{eqnarray}\label{eq:params}
    \vec{\theta}_{\mathrm{GR}} &=& \biggl\{ \ln\,D_L,\,t_c,\,\phi_c,\, \ln\,\mathcal{M},\,\eta,\,\chi_1,\,\chi_2 \biggr\},\\
    \vec{\theta}_{\mathrm{NGR}} &=& \biggl\{  \{\delta\hat{\phi}_k\},\,\{\delta\hat{\phi}_{kl}\} \biggr\},
\end{eqnarray}
Here, we do not consider the effect of different configurations of extrinsic parameters like sky location, polarization, and orientation angles of the source with respect to the detector, as we are primarily interested in studying the correlations of the PN deformation parameters with the intrinsic parameters of the binary, such as its component masses and spins. The modulation on the GW waveform because of the motion of the LISA spacecrafts~\cite{Cutler98}, which depends on the localization, polarization, and orientation angles predominantly improves the luminosity distance measurement. The estimates on the intrinsic parameters remain unaffected as these angular extrinsic parameters are weakly correlated with the intrinsic parameters~\cite{ALISA06,BBW05a,Toubiana:2020vtf}. Therefore, we use sky and polarization angle averaged LISA instrumental noise PSD given in~\cite{Babak2017}. This noise PSD also factors in the fact that LISA arms make an angle of $60^{\circ}$ with each other. We sum over the two independent low-frequency channels by dividing the total LISA instrumental noise PSD in~\cite{Babak2017} by 2. Since the noise PSD already takes care of the sky location and polarization averaging factors, we do not need to consider them again in the waveform, except for averaging over orientation/inclination angles which amounts to multiplying the amplitude by $\sqrt{4/5}$. The confusion noise due to unresolved galactic binaries limits the low-frequency regime, typically $\lesssim$ 1mHz. The sum of the instrumental and the galactic confusion noise constitutes the total LISA noise PSD. The galactic confusion noise for four years of observation with LISA is taken from~\cite{Mangiagli:2020rwz}.  
\subsection{The Fisher information matrix formalism}
\label{sec:FIM}
The bounds on the parameters ($\vec{\theta}$) that characterize the GW signal buried in the noise can be obtained with the help of the Fisher information matrix (FIM) formalism when the SNR is high~\cite{Cramer46,Rao45}. When the noise is stationary and Gaussian, and within the limit of high SNR, the error in the estimation of the parameters, $\vec{\theta}$ from the GW signal follows a multivariate Gaussian distribution,
\begin{equation}
    {\mathcal P} (\Delta\theta_i ) \propto {\rm \exp}\left[ -\frac{1}{2} \Gamma_{jk}\Delta\theta^j\Delta\theta^k\right],
\label{eq:Fisher_likelihood}
\end{equation}
where, $\Gamma_{jk}$ is the FIM whose components are given by the noise-weighted inner products of the waveform derivatives with respect to $\vec{\theta}$~\cite{CF94,PW95,AISS05,Vallisneri07}:
\begin{equation}\label{eq:noise-weighted}
\Gamma_{jk}=\Big \langle\frac{\partial {\tilde h}(f)}{\partial
\theta^j}, \frac{\partial {\tilde h}(f)}{\partial \theta^k} \Big \rangle.
\end{equation}
The inner product $<\,,>$ is defined as,
\begin{equation}
    \langle a\,, b\rangle =2 \,\int_{f_{\rm low}}^{f_{\rm high}}\frac{a(f)\,b(f)^{*}+a(f)^{*}\,b(f)}{S_n(f)}\,df,
    \label{eq:int}
\end{equation}
where, $*$ denotes the complex conjugate, and $f_\mathrm{low}$ and $f_\mathrm{high}$ are the lower and upper cutoff frequencies, determined by the GW source properties, LISA's observational time period, $T_{\mathrm{obs}}$ and noise PSD, $S_n(f)$. The lower frequency cutoff is chosen such that the GW signal from the inspiraling SMBBH lasts for four years prior to its merger ($T_{\mathrm{obs}}$ = 4 yrs) but is not lower than the low-frequency limit of the LISA noise PSD, which is $10^{-4}$ Hz:
\begin{equation}
\label{flow}
f_{\mathrm{low}} = \mathrm{Max}\left[ 10^{-4}, 4.149 \times 10^{-5} \left(\frac{M_c}{10^6}\right)^{-\frac{5}{8}}\times T_{\mathrm{obs}}^{-\frac{3}{8}} \right].
\end{equation}
The upper cut-off frequency ($f_{\mathrm{high}}$) is given by the frequency $f_{\mathrm{IMR}}$ at which the ratio between the GW characteristic amplitude ($2 {\sqrt f} |\tilde h(f)|$) and the LISA noise amplitude spectral density is no less than 10\%~\cite{Datta:2020vcj, Gupta:2020lxa} but does not exceed the usual upper frequency limit of the LISA noise PSD (0.1 Hz):
\begin{equation}
\label{fhigh}
f_{\mathrm{high}} = \mathrm{Min}\left[ f_{\mathrm{IMR}}, 0.1\right].
\end{equation}
The variance-covariance matrix, ($C_{jk}$) is obtained by inverting the FIM. The square-root of the diagonal elements of $C_{jk}$ yields the 1-$\sigma$ uncertainties in the measurement of the parameters $\vec{\theta}$.
\subsection{Construction of the best-measured linear combinations using principal component analysis}
\label{sec:PCAonFisher}
As discussed earlier, multiparameter tests of GR are rigorous, strong-field, null tests of GR, where multiple PN deformations are simultaneously measured along with the GR parameters of the binary~\cite{AIQS06a,Gupta:2020lxa,Datta:2020vcj}.
However, for a full 8-parameters test, where all the eight PN deformation parameters ($\vec{\theta}_{\mathrm{NGR}}$) are simultaneously measured with the seven GR parameters ($\vec{\theta}_{\mathrm{GR}}$), the correlations between the PN deformation parameters make the corresponding fifteen-dimensional FIM ill-conditioned and hence non-invertible~\cite{AIQS06a}.

The correlations can be eliminated by using the method of PCA on the marginalized (over $\vec{\theta}_{\mathrm{GR}}$) FIM corresponding to the eight PN deformation parameters, $\tilde\Gamma$, computed using the Schur complement method~\cite{Datta:2020vcj,Gupta:2020lxa,Datta:2022izc}. 
We first diagonalize the marginalized FIM:
\begin{equation}\label{eq:PCA-decomposition}
\tilde\Gamma = U\,S\,U^{T},
\end{equation}
where, $S \equiv {\rm diag}\{\lambda_1, \lambda_2,..,\lambda_n\}$ is the diagonal matrix with $n$ being the dimension of $\tilde\Gamma$ or the number of PN deformation parameters simultaneously measured. The elements of $S$ are the eigenvalues of the system arranged in decreasing order, and $U$ is the transformation matrix whose columns are the eigenvectors orthogonal to each other. Upon plugging Eq.~(\ref{eq:PCA-decomposition}) into Eq.~(\ref{eq:Fisher_likelihood}), we can clearly see that Eq.~(\ref{eq:PCA-decomposition}) defines a linear transformation of the old correlated basis ($\vec{\theta}_{\mathrm{NGR}}$) to a new orthogonal basis, in which the FIM is diagonal:
\begin{equation}\label{eq:linearcomb}
\delta\hat{\phi}^{(i)}_{\mathrm{PCA}} = \sum_k U^{ik} \delta\hat{\phi}_k.
\end{equation}
The new set of deformation parameters, $\delta\hat{\phi}^{(i)}_{\mathrm{PCA}}$ defines the new orthogonal basis,  which are the optimal linear combinations of the original PN deformation parameters as shown in Eq.~(\ref{eq:linearcomb}). We will call these new deformation parameters as PCA parameters in the rest of the paper. The magnitude of the eigenvalues decides the dominance of the associated PCA parameters. For instance, the eigenvector ${U}^{1k}$ associated with the highest eigenvalue gives the most dominant PCA parameter, $\PCAone$.
Furthermore, the eigenvectors, ${U}^{ik}$, are uniquely determined by the source properties of the GW event. The diagonal matrix $S$ can be easily inverted to obtain the covariance matrix for the new PCA parameters.
The 1$\sigma$ errors in the measurement of the PCA parameters are given by,
\begin{equation}
\Delta[\delta\hat{\phi}^{(i)}_{\mathrm{PCA}}] =  \sqrt{\frac{1}{\lambda^{i}}}.
\label{eq:PCAerror}
\end{equation}
For a given GW source, the trace of the marginalized FIM, $\tilde\Gamma$ is a measure of the total information it carries and remains invariant under the linear transformation shown in Eq.~(\ref{eq:PCA-decomposition}). Therefore, due to the transformation in Eq.~(\ref{eq:PCA-decomposition}), the total information carried by $\tilde\Gamma$ gets optimally redistributed among the new PCA parameters. The magnitude of the eigenvalues of the corresponding PCA parameters determines the amount of information they carry. 

Naturally, the most dominant PCA parameter, $\PCAone$, has the most information about the dynamics of the source. The PCA parameters with the lowest eigenvalues have the least information about the system and are responsible for the singular behavior of the covariance matrix. Therefore, such sub-dominant PCA parameters could be dropped, depending on a pre-decided threshold on the information to be retained by the PCA parameters.

The proportion of information in a single PCA parameter relative to the total information in all the PCA parameters is defined as~\cite{Datta:2022izc},
\begin{equation}\label{eq:POV}
I_k = \frac{\lambda_k}{\sum_{k=1}^{n} \lambda_k},
\end{equation}
where, $\lambda_k\,(k=1, \ldots,n$) are the eigenvalues obtained by performing the diagonalization shown in Eq.~(\ref{eq:PCA-decomposition}) to the $n$-dimensional marginalized FIM obtained from an $n$-parameter test.
The cumulative information carried by the $m$ dominant PCA parameters is calculated as:
\begin{equation}\label{eq:cumPOV}
    \sum_{k=1}^{m} I_k = \frac{\sum_{k=1}^{m} \lambda_k}{\sum_{k=1}^{n} \lambda_k}\,\,\,\,\,(m\leq n).
\end{equation}
We will use Eq.~(\ref{eq:cumPOV}) to quantify the number of PCA parameters required to capture the information carried by the FIM corresponding to the original PN deformation parameters up to a given threshold. 
\section{Measurement uncertainties on the PCA parameters}\label{sec:PCAerrors}
\begin{figure}[htp]
 \begin{minipage}{.48\textwidth}
  \includegraphics[width=1\textwidth]{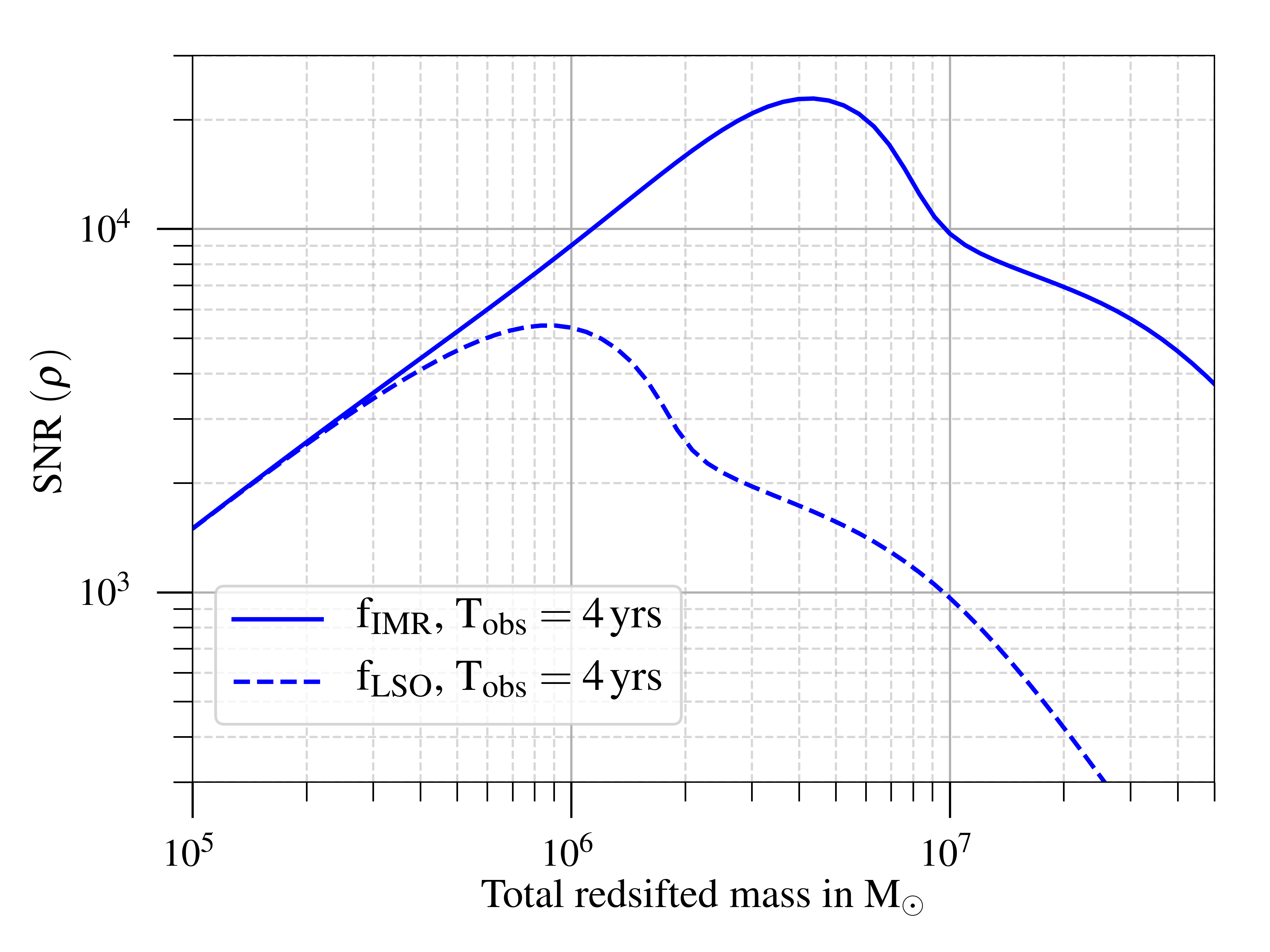}
  \caption{Signal to noise ratio as a function of total redshifted mass, calculated for four years of observation time with LISA. The solid line indicates the SNRs for the SMBBH systems calculated using the full \textsc{IMRPhenomD} waveform, with upper cutoff frequency given in Eq.~(\ref{fhigh}). The dashed line indicates the SNRs calculated till the last stable orbit frequency given by, $f_{\mathrm{LSO}} = 1/(6^{3/2}\pi M_z)$.}
  \label{fig:SNRs}
 \end{minipage}
 \begin{minipage}{.48\textwidth}
  \includegraphics[width=1\textwidth]{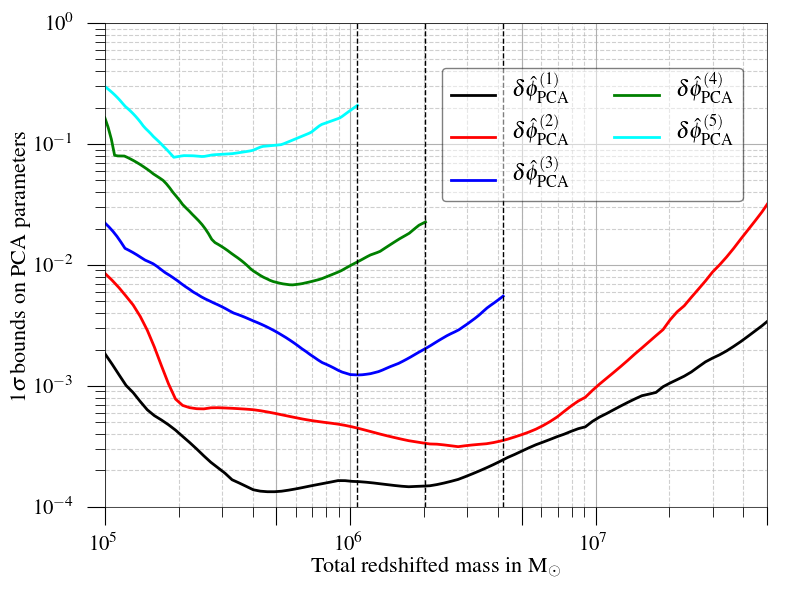}
  \caption{1-$\sigma$ bounds on the PCA parameters as a function of total redshifted mass. The vertical dashed lines indicates the maximum mass up to which a PCA parameter can be measured. All the sources are placed at a luminosity distance of 3 Gpc.}
  \label{fig:PCAerrors}
 \end{minipage}
\end{figure}
We use the method described in Sec.~\ref{sec:PCAmethod} to calculate the bounds on the PCA parameters as a function of the total redshifted mass. We choose the redshifted masses in the range of $10^5$--$5\times10^7$ $\msun$, placed at a luminosity distance of 3 Gpc.
Such sources will be visible in the LISA band for years before they merge and will accumulate substantial SNR [$\mathcal{O} (10^3)$--$\mathcal{O} (10^4)$] during their long inspiral phase. Figure~\ref{fig:SNRs} (dashed curve) shows that SNR in the inspiral regime is largest for SMBBHs with masses between $2\times 10^5$--$5\times10^6$ $\msun$ and therefore should be ideal for this inspiral-based analysis.
We avoid zero spin magnitudes ($\chi_1=\chi_2=0$) or equal mass binary ($q=1$) configurations as they make certain terms associated with spin-orbit and spin-spin couplings in the PN coefficients starting from 1.5PN to 3.5PN order vanish.
Therefore, we choose the dimensionless spin magnitudes of the more massive ($m_1$) and the less massive ($m_2$) component to be $\chi_1 = 0.3$ and $\chi_2=0.2$ respectively, and the mass ratio is taken to be $q=2$. In Appendix~\ref{sec:PCA_error_qandchi}, we explore a wider parameter space by varying the mass ratio and spins of the sources.

Figure.~\ref{fig:PCAerrors} shows the 1-$\sigma$ bounds on the first five dominant PCA parameters are constrained to $\lesssim 0.2$. The best bounds are of the order of $10^{-4}$ for the first and second most dominant PCA parameter, $\PCAone$ and $\PCAtwo$, $10^{-3}$ for the third and fourth dominant PCA parameter, $\PCAthree$ and $\PCAfour$, and $10^{-2}$ for the fifth dominant PCA parameter $\PCAfive$. These best bounds are obtained for masses between $2\times 10^5$--$5\times10^6$ $\msun$ because
they have a large number of GW cycles and the most SNR in the inspiral regime as discussed in the previous paragraph. Furthermore, the bounds on the PCA parameters worsen monotonically from the dominant to the more sub-dominant ones. This happens because the sub-dominant PCA parameters are associated with smaller eigenvalues of $\tilde{\Gamma}$, and the error in their measurement is inversely proportional to the square root of their eigenvalues as shown in Eq.~(\ref{eq:PCAerror}). Figure.~\ref{fig:PCAerrors} also tells that the number of PCA parameters that are measurable decreases with the increasing total mass of the system, due to the drop in the number of inspiral-cycles and inspiral-SNR with increasing mass.

\begin{figure}[htp]
\centering
\includegraphics[width=0.49\textwidth]{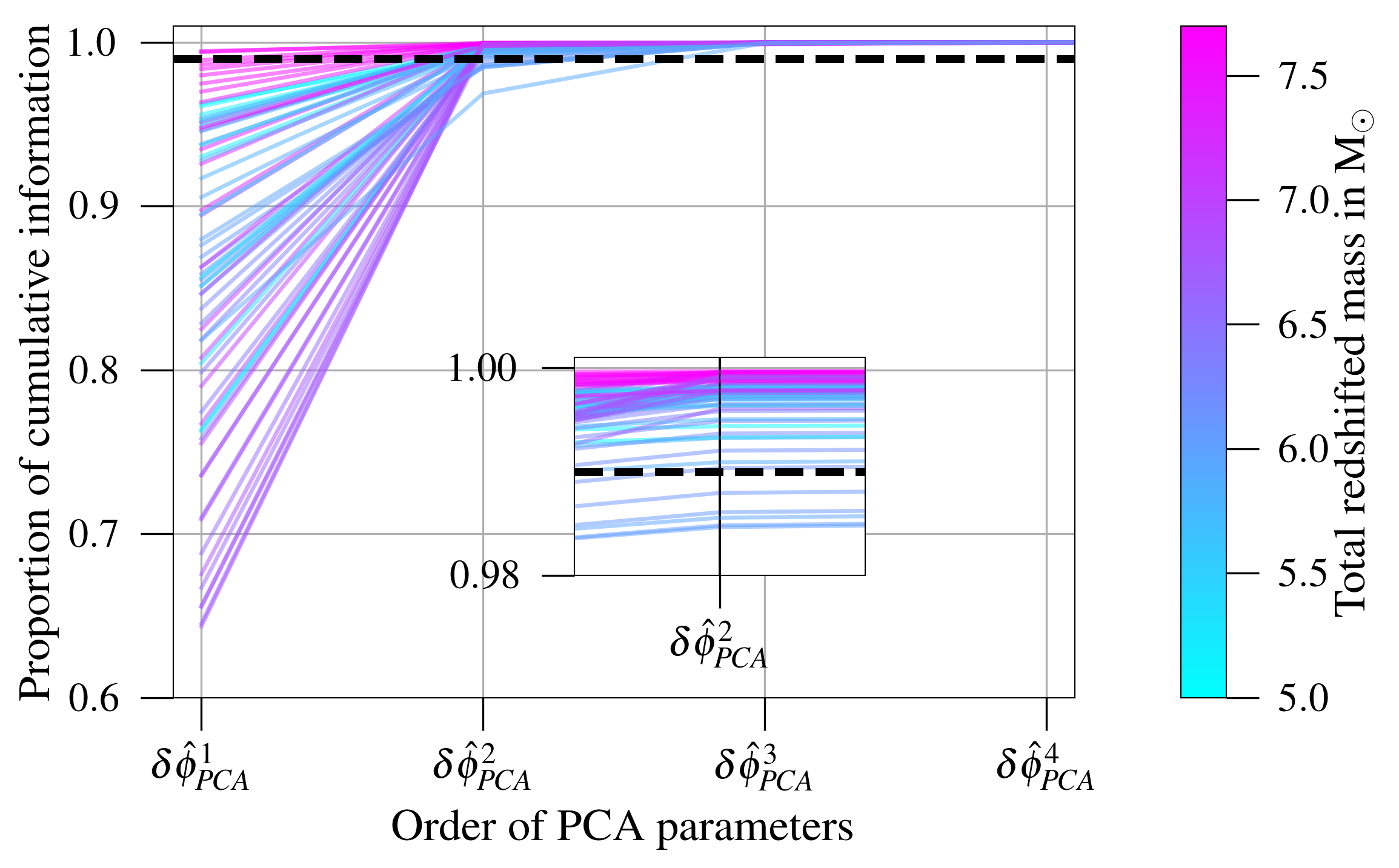}
\captionsetup{justification=raggedright,singlelinecheck=false}
\caption{The cumulative information across the PCA parameters as a function of total redshifted mass of the system. The black dashed horizontal line mark the threshold which is 99\% of the total information. The colour bar shows the masses in logarithmic scale of base 10.}
\label{fig:cumPOI}
\end{figure}

The hierarchy in the measurement of the PCA parameters can also be interpreted in terms of the amount of information they capture. The greater the eigenvalue of the PCA parameter, the greater the information it carries and the better its measurement. We calculate the normalized cumulative information that each PCA parameter carries using Eq.~(\ref{eq:cumPOV}) and plot them in Fig.~\ref{fig:cumPOI} as a function of total mass depicted by the colour bar. The saturation in the cumulative information when the sub-leading PCA parameters beyond $\PCAtwo$ are added is due to the gradual decline in the proportion of information and, therefore, the measurability of the sub-leading PCA parameters.

It is essential to identify the sub-dominant PCA parameters carrying negligible information. They are responsible for rendering an ill-conditioned covariance matrix and are prone to be noise-dominated for real gravitational wave signals~\cite{Saleem:2021iwi,Datta:2022izc}. The insignificant subdominant PCA parameters can be safely neglected depending upon a predetermined threshold on the total information the PCA parameters should retain.
We set this threshold at 99\% and determine the number of PCA parameters that carry at least 99\% of the information.

Figure.~\ref{fig:cumPOI} clearly shows that the most dominant PCA parameter for the highest and lowest masses carries more than 90\% of the information. Still, none cross the 99\% mark and require the second-dominant PCA parameter to meet the threshold value. However, the contribution from the second dominant PCA parameter is much smaller than the dominant one. 
For masses between $2\times 10^5$--$5\times10^6$ $\msun$, the second-dominant PCA parameter carries a significant amount of information, playing a much more crucial role in meeting the required criterion. The second-dominant PCA parameter for a few SMBBH systems of $\sim 10^6$ $\msun$ cannot meet the 99\% threshold, which suggests that such systems require the third-dominant PCA parameters to satisfy the criterion. This is expected as they are the most inspiral-dominated systems as seen in Fig.~\ref{fig:SNRs}.

Constraining as many of the sub-dominant PCA parameters as possible with good precision is important, because they capture the late-time dynamics in the phase evolution. This is why masses between $2\times 10^5$--$5\times10^6$ $\msun$ are ideal for this test with LISA, as we can put strong bounds on the most number of PCA parameters. We explore this feature of the subdominant PCA parameters in greater detail in the next section and Appendix~\ref{sec:PCAcomp}. In the next section, we will make a detailed comparison of the measurement of the PCA parameters between LISA and the next-generation ground-based detectors like CE and ET.
%
\section{Comparison in the measurement of the PCA parameters for LISA and ground based detectors}\label{sec:PCAcomparison}
In a recent paper~\cite{Datta:2022izc}, we computed the PCA parameters with next-generation ground-based detectors like CE and ET. We studied the properties of the best-measured PCA parameter as a function of the total redshifted mass ranging from 20-200$M_{\odot}$. In this section, we will draw a qualitative comparison between the PCA parameters computed using LISA, CE, and ET for systems with the highest inspiral SNR in their respective bandwidths and placed at luminosity distances which are representatives of the corresponding detector's typical reach. Therefore, we consider systems with a total redshifted mass of 200 $M_{\odot}$ and 450 $M_{\odot}$ at a luminosity distance of 500 Mpc in the CE and ET band respectively (SNR of ~1920 and ~1520 respectively). For LISA, we choose a system with a total redshifted mass of $8.71\times 10^5$ $M_{\odot}$ at a luminosity distance of 3 Gpc (SNR of ~5425). We deliberately fix the luminosity distance, not the SNR, as we intend to mimic a realistic observing scenario and exploit the higher SNRs for LISA compared to CE or ET for their respective chosen systems to reflect on the bounds. The mass ratio for all the systems is chosen to be $q=2$, and the individual spin magnitudes are fixed at $\chi_1 = 0.3$ and $\chi_2=0.2$.
Using these representative systems, we compute their PCA parameter using the method described in Sec.~\ref{sec:PCAonFisher}. 
We compare these best-measured PCA parameters computed for different detectors in terms of their measurability and composition.
\begin{table}[htbp]
\begin{tabular}{|M{2cm}||M{2.2cm}|M{1.5cm}|M{1.5cm}|}
 \hline
 & \rule{0pt}{5mm}LISA  & CE & ET \\ [0.2cm]
 \cline{2-4}
       PCA parameters & \rule{0pt}{5mm}$8.71\times 10^5\,M_{\odot}$ (5425.5) & 200 $M_{\odot}$ (1921.6) & 450 $M_{\odot}$ (1519.4) \\ [0.2cm]
 \hline
 \hline
 \vspace{0.2cm}
 $\mathbf{\PCAone}$   & $\mathbf{1.61\times 10^{-4}}$ & $\mathbf{9.4\times 10^{-4}}$  & $\mathbf{1.1\times 10^{-3}}$\\ [0.2cm]
 $\mathbf{\PCAtwo}$  & $\mathbf{4.86\times 10^{-4}}$  & $\mathbf{3.1\times 10^{-3}}$  & $\mathbf{2.5\times 10^{-3}}$\\ [0.2cm]
 $\mathbf{\PCAthree}$ & $\mathbf{1.35\times 10^{-3}}$  & 0.08 & 0.07\\ [0.2cm]
 $\PCAfour$  & $8.57\times 10^{-3}$ &  &  \\ [0.2cm]
 $\PCAfive$ & 0.12 &    & \\ [0.2cm]
 \hline
 \end{tabular}
 \caption{The 1-$\sigma$ bounds on the PCA parameters computed with LISA, CE and ET for systems with detector frame masses of $8.71\times 10^5\,M_{\odot}$, 200 $M_{\odot}$ and 450 $M_{\odot}$ and SNRs of 5425.5, 1921.6 and 1519.4, respectively. The SNRs are calculated up to the last stable orbit frequency, $f_{\mathrm{LSO}} = 1/(6^{3/2}\pi M_z)$. }
\label{tab:comparison-pca}
\end{table}
\begin{table}[htbp]
\centering
 \begin{tabular}{|M{2.1cm}||M{0.9cm}|M{1.2cm}|M{1.2cm}|M{1.5cm}|}
 \hline
 \multicolumn{1}{| c ||}{}  & \multicolumn{4}{ c |}{Cumulative percentage of information}\\ [0.2cm]
 \cline{2-5}
 Events & \rule{0pt}{5mm}$\PCAone$ & $\PCAtwo$ & $\PCAthree$ & $\PCAfour$ \\ [0.2cm]
 \hline
 \hline
 \vspace{0.2cm}
LISA ($8.71\times 10^5\,M_{\odot}$) & 88.92 & 98.7 & 99.96 & 99.99 \\ [0.2cm] 
CE (200 $M_{\odot}$) & 91.34 & 99.982 & 99.99 & 100\\ [0.2cm] 
ET ($450\,M_{\odot}$) & 84.423 & 99.981 & 99.99 & 100\\ [0.2cm] \hline
 \end{tabular}
 \caption{Cumulative percentage of information carried by the four most dominant PCA parameters computed with LISA, CE and ET for systems with total redshifted masses of $8.71\times 10^5\,M_{\odot}$, 200 $M_{\odot}$ and 450 $M_{\odot}$ respectively. }
\label{tab:comparison-pov}
\end{table}
\begin{table*}
       \begin{center}
              \begin{tabular}{| M{2.2cm} | M{2.5cm} | M{1cm} | M{1cm} |  M{1cm} | M{1cm} | M{1cm} | M{1cm} | M{1cm} | M{1cm} |}
              \hline
              Event & PCA parameters & \rule{0pt}{4mm}\boldmath{$|\mathrm{U}_{i\,0}|$} & \boldmath{$|\mathrm{U}_{i\,2}|$} & \boldmath{$|\mathrm{U}_{i\,3}|$} & \boldmath{$|\mathrm{U}_{i\,4}|$} & \boldmath{$|\mathrm{U}_{i\,5l}|$} & \boldmath{$|\mathrm{U}_{i\,6}|$} & \boldmath{$|\mathrm{U}_{i\,6l}|$} & \boldmath{$|\mathrm{U}_{i\,7}|$} \\ [0.1cm]
              \hline\hline 
              \vspace{0.2cm}
              \multirow{4}{2cm}{LISA ($8.71\times 10^5\,M_{\odot}$)}& $\PCAone$ & 0.52 & 0.61 & 0.593 & 0.053 & 0.048 & 0.0002 & 0.005 & 0.001 \\ [0.2cm] 
                    & $\PCAtwo$ & 0.847 & 0.444 & 0.288 & 0.004 & 0.035 & 0.001 & 0.021 & 0.007 \\ [0.2cm] 
                    & $\PCAthree$ & 0.107 & 0.517 & 0.563 & 0.221 & 0.554 & 0.011 & 0.205 & 0.077  \\[0.2cm] 
                    & $\PCAfour$ & 0.01 & 0.391 & 0.461 & 0.069 & 0.606 & 0.019 & 0.481 & 0.176 \\ [0.2cm]
                    & $\PCAfive$ & 0.002 & 0.097 & 0.176 & 0.462 & 0.378 & 0.074 & 0.63 & 0.448 \\ [0.2cm]
              \hline
              \vspace{0.2cm}
              \multirow{3}{2cm}{CE (200 $M_{\odot}$)}& $\PCAone$ & 0.887 & 0.284 & 0.352 & 0.046 & 0.075 & 0.001 & 0.017 & 0.007 \\ [0.2cm] 
                    & $\PCAtwo$ & 0.445 & 0.355 & 0.748 & 0.139 & 0.292 & 0.005 & 0.102 & 0.039\\ [0.2cm] 
                    & $\PCAthree$ & 0.053 & 0.17 & 0.075 & 0.023 & 0.125 & 0.004 & 0.972 & 0.038 \\[0.2cm] 
                \hline
                \vspace{0.2cm}
              \multirow{2}{2cm}{ET (450 $M_{\odot}$)}& $\PCAone$ & 0.871 & 0.297 & 0.378 & 0.051 & 0.083 & 0.001 & 0.02 & 0.008\\[0.2cm] 
                    & $\PCAtwo$ & 0.478 & 0.364 & 0.735 & 0.132 & 0.268 & 0.005 & 0.088 & 0.034  \\[0.2cm] 
                    & $\PCAthree$ & 0.102 & 0.577 & 0.024 & 0.176 & 0.664 & 0.017 & 0.402 & 0.15 \\[0.2cm] 
            \hline                                
             \end{tabular}
      \end{center}
\caption{Magnitudes of the elements of the eigenvectors, $\mathrm{U}_{ik}$ which constitute the corresponding PCA parameters, $\delta\hat{\phi}^{(i)}_{\mathrm{PCA}}$ for each of the binary black hole systems considered.}
\label{tab:comparison-coeff}
\end{table*}

Table~\ref{tab:comparison-pca} shows the 1-$\sigma$ bounds on the PCA parameters for LISA, CE, and ET. The numbers in bold font represent the PCA parameters which carry at least 99.9\% of the total information. Table~\ref{tab:comparison-pov} shows that capturing 99.9\% of the total information requires the first three dominant PCA parameters for LISA, whereas the first two dominant PCA parameters are sufficient for CE and ET. 
LISA yields the best bounds on the PCA parameters compared to CE/ET because of the larger number of inspiral cycles and higher inspiral SNR due to the longer duration of the inspiral signal in its band.

Table~\ref{tab:comparison-coeff} shows the values of the elements of the transformation matrix, $U^{ik}$. The columns of $U^{ik}$ are the eigenvectors that constitute the PCA parameters as shown in Eq.~(\ref{eq:linearcomb}). We have shown the composition of only those PCA parameters, which can be measured with a 1-$\sigma$ error of $\leq \mathcal{O}10^{-2}$.
The magnitude of the elements of $U^{ik}$ determines the weight with which the associated PN deformation parameter contributes to the errors on the PCA parameter. Hence, by studying the composition of the best-measured linear combinations, we can identify the PN deformation parameters that the PCA parameters are most sensitive to.

From Table~\ref{tab:comparison-coeff}, it is clear that the two leading PCA parameters, $\PCAone$, and $\PCAtwo$ for LISA, mostly probe the PN deformation parameters at 0PN, 1PN, and 1.5PN. Whereas the third dominant PCA parameter $\PCAthree$
shows significant contribution from the deformation parameter at 2.5PN-log and reduced contribution from 0PN compared to  $\PCAone$ and $\PCAtwo$. Similarly, the other sub-dominant PCA parameters, $\PCAfour$, and $\PCAfive$, probe the deformations at higher PN orders like 2PN, 2.5PN-log, 3PN-log and 3.5PN better than the dominant PCA parameters. A similar feature is also seen for CE and ET. The leading PCA parameter is dominated by the deformation parameter at 0PN and the second-leading PCA parameter by the deformation parameters at 0PN, 1PN and 1.5PN. The third-leading PCA parameter has significant contributions from the deformation parameters at 2.5PN-log and 3PN-log. However, the third-leading PCA parameter for LISA has better bounds than CE and ET by three orders of magnitude. 

Therefore, the numbers in Table~\ref{tab:comparison-coeff} and Table~\ref{tab:comparison-pca} tell us that the sub-dominant PCA parameters are sensitive to the deformation parameters at higher PN orders. LISA can measure the first five dominant PCA parameters with a 1-$\sigma$ uncertainty of $\leq 0.2$, better probing the higher PN order deformations than next-generation terrestrial detectors. Furthermore, the $U^{ik}$ coefficients depend on the intrinsic source properties. In Appendix~\ref{sec:PCAcomp}, we show the variation in the magnitude of the coefficients as a function of the total mass of the SMBBH systems.

\section{Conclusions}
\label{sec:conclusion}
We discussed the prospect of performing multiparameter tests of GR with LISA using SMBBHs, which are the ideal GW sources for this test. The full multiparameter test of GR, where all the PN deformation parameters are simultaneously measured along with the GR parameters, is challenging to perform because of the correlations between the PN deformation parameters in terms of the intrinsic properties of the system. We used principal component analysis to construct an orthogonal set of new deformation parameters called PCA parameters. The PCA parameters are the best-measured optimal linear combinations of the original PN deformation parameters. Therefore, they are sensitive to the overall structure of the GW phase evolution as predicted by PN approximation to GR.

We studied the variation in the bounds of the PCA parameters as a function of the system's total mass. We obtain the best bounds for the most inspiral-dominated sources with masses between $2\times 10^5$--$5\times10^6$ $\msun$. The number of measurable PCA parameters with significant information decreases with the system's increasing mass. With an SMBBH system of $\sim 7\times10^5\,M_{\odot}$, we can measure up to five most dominant PCA parameters, with a 1-$\sigma$ uncertainty of $\lesssim 0.2$. 
By looking into the proportion of information carried by the PCA parameters, we find that SMBBH systems in the mass range $2\times 10^5$--$5\times10^6$ $\msun$ will require at least three most dominant PCA parameters to capture at least 99\% of the total information. 

We also compare the measurability1 of the PCA parameters computed with LISA, CE, and ET. We choose sources with the highest inspiral SNRs for each of the detectors and hence are ideal for this PCA-assisted multiparameter tests of GR. We find that LISA can measure more number of PCA parameters (five) with 1-$\sigma$ of $\leq0.2$, than CE (two) and ET (two). The composition of the sub-leading PCA parameters suggests that LISA facilitates a better probe to the higher PN orders than the next-generation terrestrial detectors CE and ET.

In this study, we only consider SMBBH systems with zero eccentricity. We also consider the orbital angular momentum vector to be aligned with the component spin vectors, showing zero precession and higher order modes (beyond the leading quadrupolar mode). However, SMBBH systems, four years before their merger, may have needed more time to circularize and therefore have non-negligible orbital eccentricity. 
Non-zero eccentricity, precession, and higher order modes in the system cause faster loss of orbital energy by gravitational waves over different time scales, inducing extra characteristic modulations in the amplitude of the radiated gravitational waves. Therefore, including these features may improve the parameter estimation, resulting in better bounds in the PCA parameters~\cite{MAIS10,ChrisAnand06b,AISSV07,SW09,HKJ11,Favata:2021vhw}. We plan to study the effect of eccentricity, precession, and higher order modes in the measurability of the PCA parameters and quantify the systematic biases due to their ignorance in a future work.


\acknowledgments
We thank  K. G. Arun, Anuradha Gupta, B. S Sathyaprakash, Pankaj Saini and Sajad A. Bhat for providing useful comments on the manuscript. We are also thankful to Arnab Dhani, Nathan K. Johnson-McDaniel and Parthapratim Mahapatra for very useful discussions.
We thank Sebastian Khan for sharing his Mathematica code of the \textsc{IMRPhenomD} waveform model with us. S.D. was partially supported by the Swarnajayanti grant DST/SJF/PSA-01/2017-18 of the Department of Science and Technology, India. S.D also acknowledge support from Infosys Foundation. We use the following software packages for the computations in this work: \textsc{NumPy} \cite{vanderWalt:2011bqk}, \textsc{SciPy} \cite{Virtanen:2019joe}, \textsc{Matplotlib} \cite{Hunter:2007}. 

\appendix
\section{Composition of the PCA parameters}
\label{sec:PCAcomp}
In this appendix, we will identify the PN deformation parameters that the PCA parameters probe by studying the composition of the linear combinations. Specifically, we will look into the elements of the transformation matrix, $\mathrm{U}^{ik}$ shown in Eq.~(\ref{eq:linearcomb}), whose columns are the orthogonal eigenvectors. These eigenvectors build the PCA parameters, with the elements as coefficients determining the contribution from the associated PN deformation parameter. The eigenvectors are unique for every source binary. 
Therefore, we study the variation in the magnitude of the coefficients for three supermassive binary black hole systems with masses $2\times 10^5\,\msun$, $10^6\,\msun$ and $10^7\,\msun$.

Table~\ref{tab:coeffs} shows that the coefficients of the first three PN deformation parameters, 
$\dphizero$, $\dphitwo$,  and $\dphithree$ contribute the most to the first and third most dominant PCA parameters, $\PCAone$ and $\PCAthree$. 
The 0PN and 1PN deformation parameters, $\dphizero$ and $\dphitwo$ contribute the most to the second dominant PCA parameter for $2\times 10^5\,M_{\odot}$. However, the magnitude of the coefficient corresponding to the 1.5PN  deformation parameter rises with increasing total mass.
The contribution of the coefficients of PN deformation parameters at 2PN, 2.5PN-log, and 3PN-log to the third dominant PCA parameter also increases with the system's total mass. The higher PN orders carry information about the late time dynamics, constituting the sub-leading effects, and are more pronounced for heavier SMBBHs. Therefore, the third dominant PCA parameter tends to capture these effects better with increasing mass. 

The coefficients of the fourth dominant PCA parameter, $\PCAfour$ show that it is dominated by the 2.5PN-log deformation parameter, $\dphifivel$ for almost all the three masses, followed by 3PN-log and 2PN deformation parameter. The 0PN deformation parameter is highly suppressed for the third and the fourth dominant PCA parameters. 

In a nutshell, Table~\ref{tab:coeffs} suggests that the first three dominant PCA parameters probe the leading order inspiral dynamics of the system, which is mainly carried by 0PN, 1PN and 1.5PN deformation parameters. The sub-dominant PCA parameters probe the late-time dynamics represented by the higher PN order deformation parameters. Hence, the sub-dominant PCA parameters for SMBBHs will help us probe higher-order strong-field effects as they are also constrained to excellent accuracy by LISA, as shown in Fig.~\ref{fig:PCAerrors}.

\begin{table*}
       \begin{center}
              \begin{tabular}{| M{2.2cm} | M{2.5cm} | M{1cm} | M{1cm} |  M{1cm} | M{1cm} | M{1cm} | M{1cm} | M{1cm} | M{1cm} |}
              \hline
              Event & PCA parameters & \rule{0pt}{4mm}\boldmath{$|\mathrm{U}_{i\,0}|$} & \boldmath{$|\mathrm{U}_{i\,2}|$} & \boldmath{$|\mathrm{U}_{i\,3}|$} & \boldmath{$|\mathrm{U}_{i\,4}|$} & \boldmath{$|\mathrm{U}_{i\,5l}|$} & \boldmath{$|\mathrm{U}_{i\,6}|$} & \boldmath{$|\mathrm{U}_{i\,6l}|$} & \boldmath{$|\mathrm{U}_{i\,7}|$} \\ [0.1cm]
              \hline\hline 
              \vspace{0.2cm}
              \multirow{4}{2cm}{$2\times 10^5\,M_{\odot}$}& $\PCAone$ & 0.549 & 0.544 & 0.631 & 0.052 & 0.021 & 0.001 & 0.025 & 0.008 \\ [0.2cm] 
                    & $\PCAtwo$ & 0.702 & 0.707 & 0.006 & 0.044 & 0.074 & 0.0007 & 0.0007 & 0.0003 \\ [0.2cm] 
                    & $\PCAthree$ & 0.45 & 0.446 & 0.766 & 0.087 & 0.016 & 0.002 & 0.055 & 0.018  \\[0.2cm] 
                    & $\PCAfour$ & 0.06 & 0.051 & 0.03 & 0.272 & 0.895 & 0.017 & 0.323 & 0.116 \\ [0.2cm]
                    & $\PCAfive$ & 0.01 & 0.038 & 0.097 & 0.36 & 0.229 & 0.019 & 0.841 & 0.316 \\ [0.2cm]
              \hline
              \vspace{0.2cm}
              \multirow{3}{2cm}{$10^6\,M_{\odot}$}& $\PCAone$ & 0.552 & 0.573 & 0.599 & 0.06 & 0.069 & 0.001 & 0.003 & 0.002 \\ [0.2cm] 
                    & $\PCAtwo$ & 0.833 & 0.413 & 0.368 & 0.028 & 0.018 & 0.00002 & 0.004 & 0.001\\ [0.2cm] 
                    & $\PCAthree$ & 0.044 & 0.576 & 0.504 & 0.218 & 0.561 & 0.011 & 0.211 & 0.078 \\[0.2cm] 
                    & $\PCAfour$ & 0.005 & 0.399 & 0.463 & 0.053 & 0.597 & 0.019 & 0.486 & 0.176 \\[0.2cm]
                    & $\PCAfive$ & 0.001 & 0.033 & 0.029 & 0.588 & 0.533 & 0.032 & 0.57 & 0.207  \\[0.2cm]
                \hline
                \vspace{0.2cm}
              \multirow{2}{2cm}{$10^7\,M_{\odot}$}& $\PCAone$ & 0.922 & 0.248 & 0.29 & 0.036 & 0.054 & 0.001 & 0.009 & 0.003\\[0.2cm]
                    & $\PCAtwo$ & 0.38 & 0.459 & 0.758 & 0.122 & 0.225 & 0.004 & 0.064 & 0.025 \\[0.2cm] 
                    & $\PCAthree$ & 0.014 & 0.531 & 0.553 & 0.279 & 0.559 & 0.006 & 0.138 & 0.046 \\[0.2cm] 
                    & $\PCAfour$ & 0.068 & 0.63 & 0.178 & 0.162 & 0.576 & 0.025 & 0.423 & 0.171 \\[0.2cm]
                    & $\PCAfive$ & 0.015 & 0.214 & 0.065 & 0.291 & 0.346 & 0.04 & 0.785 & 0.355 \\[0.2cm]
            \hline                                
             \end{tabular}
      \end{center}
\caption{Magnitudes of the elements of the eigenvectors, $\mathrm{U}_{ik}$ which constitute the corresponding PCA parameters, $\delta\hat{\phi}^{(i)}_{\mathrm{PCA}}$ for each of the SMBBH systems considered.}
\label{tab:coeffs}
\end{table*}
\section{Variation in the PCA bounds with changing mass ratio and spins}
\label{sec:PCA_error_qandchi}
\begin{figure*}
    \includegraphics[width=0.95\textwidth]{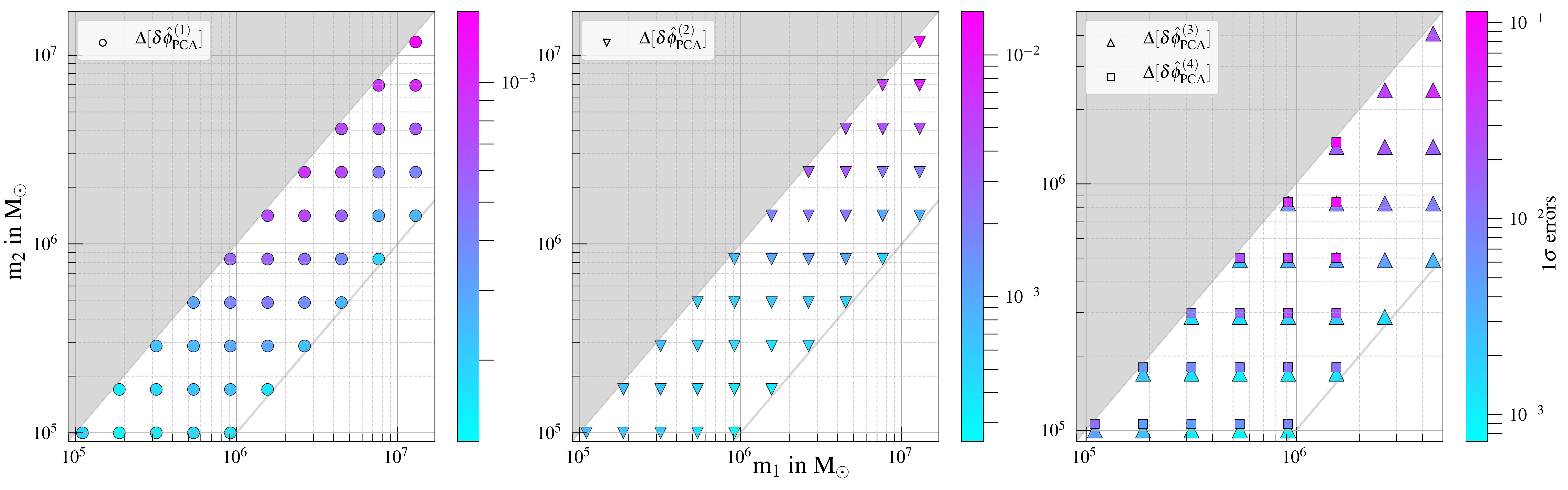} 
    \hfill
   \includegraphics[width=0.95\textwidth]{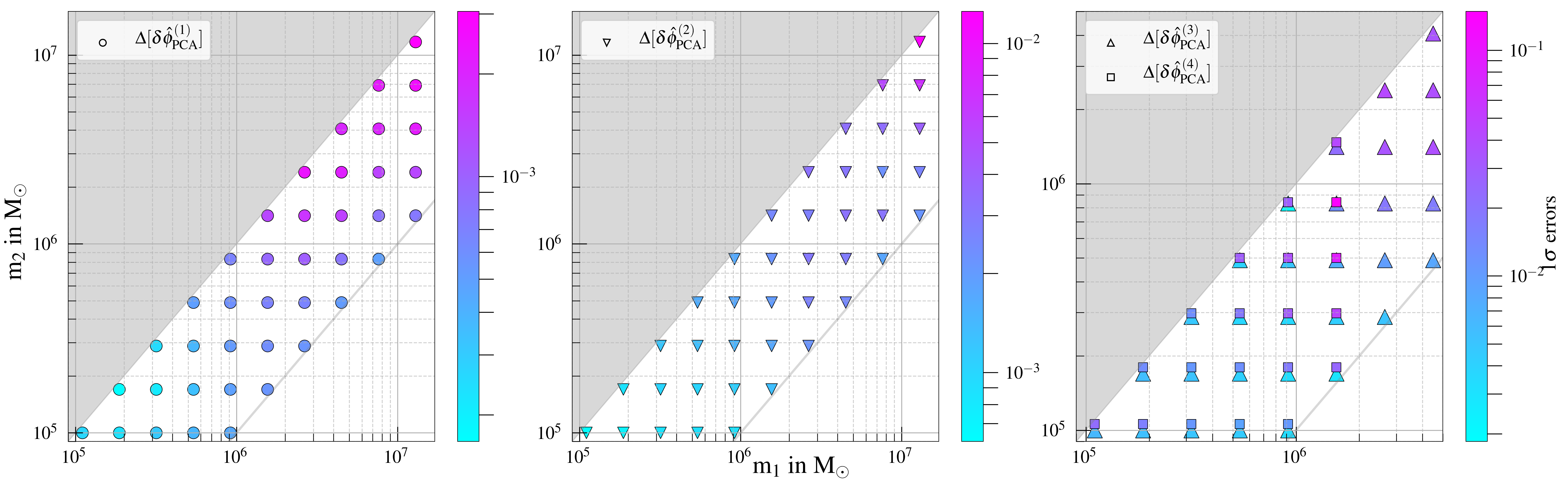}
\caption[] {\label{fig:errors_PCA_m1m2spins} Plots in the first two columns show $1\sigma$ bounds on the first and second most dominant PCA parameters, respectively. The plots in the third column show errors on the third and fourth dominant PCA parameters. The SNRs of all the systems are fixed at 3000. The plots in the top (bottom) row represent the high spin (low spin) configuration case, where the individual spin magnitudes are fixed at $\chi_1=0.9$ and $\chi_2=0.8$ ($\chi_1=0.2$ and $\chi_2=0.1$)}
\end{figure*}

The plots in Fig.~\ref{fig:errors_PCA_m1m2spins} show the variation in the $1\sigma$ bounds on the first four dominant PCA parameters for different mass ratio values and spin configurations. The systems are placed at a luminosity distance such that their total SNR is 3000. For the plots in the top (bottom) row, all the systems' individual spin magnitudes are fixed at $\chi_1=0.9$ and $\chi_2=0.8$ ($\chi_1=0.2$ and $\chi_2=0.1$). We define the mass ratio, $q$, as $m_1/m_2$ such that the primary mass is greater than the secondary mass, $m_1>m_2$. The grey shaded patch in the plots denotes the region of the parameter space where $m_1<m_2$ and is therefore disregarded. We consider systems with mass ratios between $q=1$ and $q=10$. The grey line in the plots indicates the $q=10$ limit. We also do not consider exactly equal mass systems for reasons mentioned earlier in Sec.~\ref{sec:PCAerrors}. The colour bars denote the $1\sigma$ bounds on the PCA parameters.

The mass ratio of the systems increases as we move opposite to the grey-shaded patch. The markers along each diagonal line denote systems with varying total mass but fixed mass ratio. For a fixed mass ratio, the errors in the measurement of the first two dominant PCA parameters, $\PCAone$ and $\PCAtwo$ worsen with increasing total mass because the number of inspiral cycles in the LISA band decreases. The same is not exactly true for the third and fourth dominant PCA parameters, $\PCAthree$ and $\PCAfour$, as shown in the plots in the third column of Fig.~\ref{fig:errors_PCA_m1m2spins}. They are dominated by the higher order PN deformation parameters (shown in Table.~\ref{tab:coeffs}), which come into play for heavier masses. Therefore, the best bounds on $\PCAthree$ and $\PCAfour$ are obtained for higher mass SMBBH systems compared to $\PCAone$ and $\PCAtwo$.

The scale of the colour bar attached to the top and bottom plots of the first column clearly tells that the bounds on the leading PCA parameter, $\PCAone$ are better for the high spin case than the low spin configuration. Higher spin magnitudes help break the correlation between the spins and the 1.5PN deformation parameter which has a major contribution to the bounds on the leading PCA parameter. The same applies to $\PCAthree$, although the change is negligible.

\bibliographystyle{apsrev}
\bibliography{reference.bib}

\end{document}